\documentclass[aps,prd,twocolumn,showpacs,showkeys,amsmath,amssymb,10pt]{revtex4}

\usepackage[colorlinks=true,citecolor=blue,anchorcolor=red,menucolor=red, linkcolor=red,filecolor=red,runcolor=red,urlcolor=blue,frenchlinks=true]{hyperref}
\usepackage{graphicx}
\usepackage{dcolumn}
\usepackage{txfonts}
\usepackage{bm}

\usepackage{amsfonts}
\usepackage{amsmath}
\usepackage{subfigure}
\usepackage{booktabs}
\usepackage{inputenc}
\usepackage{graphics}

\usepackage{longtable}
\usepackage{etoolbox}
\apptocmd{\sloppy}{\hbadness 10000\relax}{}{}

\usepackage{overpic}
\usepackage{indentfirst}
\usepackage{slashed}  
\usepackage{cases}
\usepackage{color}
\usepackage{multirow}
\usepackage{natbib}

\newcommand{\qq}{\langle\bar qq\rangle}
\newcommand{\ssc}{\langle \bar ss\rangle}

\newcommand{\GGb}{\langle g_s^2GG\rangle}
\newcommand{\qGqa}{\langle\bar qg_s\sigma\cdot Gq\rangle}

\newcommand{\sGsa}{\langle\bar sg_s\sigma\cdot Gs\rangle}

\usepackage{ulem}

\begin{document}

\title{Strong decays of $T^a_{c\bar s0}(2900)^{++/0}$ as a fully open-flavor tetraquark state}
\author{Ding-Kun Lian}
\author{Wei Chen}
\email{chenwei29@mail.sysu.edu.cn}
\affiliation{School of Physics, Sun Yat-Sen University, Guangzhou 510275, China}
\author{Hua-Xing Chen}
\email{hxchen@seu.edu.cn}
\affiliation{School of Physics, Southeast University, Nanjing 210094, China}
\author{Ling-Yun Dai}
\affiliation{School of Physics and Electronics, Hunan University, Changsha 410082, China} \affiliation{Hunan Provincial Key Laboratory of High-Energy Scale Physics and Applications, \\ Hunan University, Changsha 410082, China}
\author{T. G. Steele}
\email{tom.steele@usask.ca}
\affiliation{Department of Physics
and Engineering Physics, University of Saskatchewan, Saskatoon, SK, S7N 5E2, Canada}

\begin{abstract}
We have studied the strong decay properties of the recently observed $T^a_{c\bar s0}(2900)^{++/0}$ by considering it as a $cu\bar{d}\bar{s}/cd\bar{u}\bar{s}$ fully open-flavor tetraquark state with $I(J^P)=1(0^+)$. In the framework of QCD sum rules, we have calculated the three-point correlation functions of the two-body strong decay processes $T^a_{c\bar s0}(2900)^{++}\rightarrow D_s^+\pi^+$, $D^+K^+, D_s^{\ast +}\rho^+$ and $D_{s1}^+\pi^+$. The full width of $T^a_{c\bar s0}(2900)^{++/0}$ is obtained as $161.7\pm94.8$ MeV, which is consistent with the experimental observation. We predict the relative branching ratios as $\Gamma (T\rightarrow D_s\pi):
\Gamma(T\rightarrow DK):\Gamma (T\rightarrow D_s^{\ast} \rho):\Gamma (T\rightarrow D_{s1}\pi)\approx1.00:1.10:0.04:0.43$, implying that the main decay modes of $T^a_{c\bar s0}(2900)^{++/0}$ state are $D_s\pi$ and $DK$ channels in our calculations. However, the $P$-wave decay mode $D_{s1}\pi$ is also comparable and important by including the uncertainties. To further identify the nature of $T^a_{c\bar s0}(2900)^{++/0}$, we suggest confirming them in the $DK$ and $D_{s}\pi$ final states, and measuring the above ratios in future experiments. 
\end{abstract}


\maketitle

\section{Introduction}
With significant experimental and theoretical progress in the past one and a half decades, more and more charmoniumlike and bottomoniumlike states (the XYZ states) have been observed, including the hidden-charm pentaquark $P_c$ and $P_{cs}$ states, the doubly charmed $T_{cc}$ state, and the fully-charm tetraquark states~\cite{Chen:2016qju,Guo:2017jvc,Liu:2019zoy,Brambilla:2019esw,Chen:2022asf,Chen:2020uif}. These states provide a very good platform for identifying exotic states and for understanding the nonperturbative behavior of QCD. The story is ongoing with recent observation of $T^a_{c\bar s0}(2900)^{++/0}$ states \cite{LHCb:2022sfr,LHCb:2022lzp}.

Very recently, the LHCb Collaboration reported two new tetraquark candidates $T^a_{c\bar s0}(2900)^{++}$ and $T^a_{c\bar s0}(2900)^{0}$ with significance more than $9\sigma$ in the $D_s^+\pi^+$ and $D_s^+\pi^-$ invariant mass spectra of $B$ decay processes, respectively~\cite{LHCb:2022sfr,LHCb:2022lzp}. Their mass and decay width are measured respectively as 
\begin{equation}
  \begin{split}
    T^a_{c\bar s0}(2900)^{0}:\,& M=2.892\pm 0.014\pm 0.015~\text{GeV}\, ,\\
    &\Gamma=0.119\pm 0.026\pm 0.013 ~\text{GeV}\, ,\\
    T^a_{c\bar s0}(2900)^{++}:\,& M=2.921\pm 0.017\pm 0.020~\text{GeV}\, ,\\
    &\Gamma=0.137\pm 0.032\pm 0.017 ~\text{GeV}\, . \label{LHCbresults1}
  \end{split}
\end{equation}
Assuming $T^a_{c\bar s0}(2900)^{++}$ and $T^a_{c\bar s0}(2900)^{0}$ belong to the same isospin triplet, the common mass and width are determined to be
\begin{equation}
\begin{split}
M^{com} &=2.908\pm 0.011\pm 0.020~\text{GeV}\, ,
\\
\Gamma^{com} &=0.136\pm 0.023\pm 0.011~\text{GeV}\, , \label{LHCbresults2}
\end{split} 
\end{equation}
while their spin-parity was determined as $J^P=0^+$ over other possibilities by at least $7.5\sigma$. Observed in the $D_s\pi$ final states, the $T^a_{c\bar s0}(2900)^{++}$ and $T^a_{c\bar s0}(2900)^{0}$ states are composed of four different flavor quarks. They are fully open-flavor isovector $cu\bar d\bar s$ and $cd\bar u\bar s$ tetraquark states. 

The observations of $T^a_{c\bar s0}(2900)^{++}$ and $T^a_{c\bar s0}(2900)^{0}$ are in excellent agreement with 
our earlier theoretical calculations for the masses of fully open-flavor charmed $cu\bar d\bar s$ and $cd\bar u\bar s$ tetraquark states in QCD sum rules~\cite{Chen:2017rhl}. In Ref.~\cite{Chen:2017rhl}, we used the following interpolating currents with $I(J^P)=1(0^+)$ to study the fully open-flavor $sd\bar u\bar c$ and $su\bar d\bar c$ tetraquarks
\begin{align}
    J_2&=s^T_aC\gamma_\mu q_{1b}(\bar{q}_{2a}\gamma^\mu
    C\bar{c}^T_b-\bar{q}_{2b}\gamma^\mu C\bar{c}^T_a)\, , \label{currentvector0+}
    \\
    J_{5\mu\nu}&=s^T_aC\gamma_\mu q_{1b}(\bar{q}_{2a}\gamma_\nu
    C\bar{c}^T_b-\bar{q}_{2b}\gamma_\nu C\bar{c}^T_a)\, , \label{currenttensor0+}
\end{align}
in which $q_{1(2)}$ denotes up(down) quark field, and $a, b$ are color indices. These two interpolating currents are composed of a pair of axial-vector diquark and antidiquark fields ($C\gamma_\mu\times \gamma_\nu C$) in the antisymmetric $[\mathbf{\bar 3_c}]_{sq_1} \otimes [\mathbf{3_c}]_{\bar{q_2}\bar c}$ color structure. The tensor current $J_{5\mu\nu}$ can couple to the tetraquarks with quantum numbers $I(J^P)=1(0^+)$ via its traceless symmetric part $J_{5\mu\nu}(S)$ and trace part $J_{5\mu\nu}(T)$. Using these two currents, the mass sum rules in Ref.~\cite{Chen:2017rhl} gave hadron masses as $2.91\pm 0.14$ GeV ($J_2$) and $2.88\pm 0.15$ GeV ($J_{5\mu\nu}(T)$) for the $sd\bar u\bar c$/$su\bar d\bar c$ tetraquark state, which are in good agreement with the measured mass of $T^a_{c\bar s0}(2900)^{++/0}$ in Eq.~\eqref{LHCbresults2}. Furthermore, in Ref.~\cite{Chen:2017rhl} we systematically analyzed the possible decay behaviors of these fully open-flavor charmed tetraquarks. In particular, we suggested searching for the doubly-charged charmed tetraquarks in the $D_s\pi$ final states in the abstract of Ref.~\cite{Chen:2017rhl}, considering they are absolute exotic states that cannot be composed in the quark-antiquark formalism. The compact tetraquark interpretations of $T^a_{c\bar s0}(2900)^{++/0}$ are also supported by the investigations of their masses and decays in the methods of the nonrelativistic potential quark model~\cite{Liu:2022hbk} and the color flux-tube model~\cite{Wei:2022wtr}. 
These fully open-flavor charmed tetraquark states were also studied in the framework of QCD sum rules in Refs.~\cite{Chen:2016mqt,Agaev:2016lkl,Agaev:2017oay}.


Shortly after their observations, the mass of $T^a_{c\bar s0}(2900)^{++/0}$ has been reproduced by considering it as the $D_s^{\ast }\rho$ hadronic molecule in Ref.~\cite{Agaev:2022duz}. Since the mass of the $T^a_{c\bar s0}(2900)^{++/0}$ is also close to the threshold of $D^{\ast}K^{\ast}$, the $T^a_{c\bar s0}(2900)^{++/0}$ is interpreted as the $D^{\ast}K^{\ast}$ hadronic molecule in Refs.~\cite{Chen2022,Yue:2022mnf,Agaev:2022eyk}, in which both the mass and decay behaviors are studied. However, the calculations within the Bethe-Salpeter framework indicate that the $I=1$ bound state can not exist in the $D^{\ast}K^{\ast}$ system~\cite{Ke2022}, which is in conflict with the conclusions of Refs.~\cite{Chen2022,Yue:2022mnf,Agaev:2022eyk}. 
In addition, the $T^a_{c\bar s0}(2900)^{++/0}$ may also be explained as a resonance-like structure around the $D^{\ast}K^{\ast}$ threshold~\cite{Ge2022}, or as a threshold effect from the interaction of the $D^{\ast}K^{\ast}$ and $D_{s}^{\ast}\rho$ channels~\cite{Molina:2022jcd}. 

One and a half decades ago, the exotic $cq\bar s\bar q$ tetraquark state with $I(J^P)=0(0^+)$ was extensively studied to understand the structure of the charm-strange state $D_{s0}^*(2317)$~\cite{Kolomeitsev:2003ac,Maiani:2004vq,Terasaki:2003qa,Wang:2006uba}, in which the mass of such isoscalar scalar $cq\bar s\bar q$ tetraquark was given around 2.3-2.4 GeV. In Ref.~\cite{Vijande:2006hj}, the masses of the scalar $cq\bar s\bar q$ tetraquarks with $I=0$ and $I=1$ were calculated as 2731 MeV and 2699 MeV respectively. All these results are much lower than the mass of $T^a_{c\bar s0}(2900)^{++/0}$. In Ref.~\cite{Cheng:2020nho}, the authors predicted four isoscalar scalar $cu\bar s\bar d$ tetraquarks with masses (2213, 2537, 2919, 3150) MeV and decay widths (70, $>$110, 40, $>$110) MeV by using a color-magnetic interaction model, in which only the second-highest state lies close to the $T^a_{c\bar s0}(2900)^{++}$ with much narrower decay width.

To identify the nature of $T^a_{c\bar s0}(2900)^{++}$, we investigate its hadronic decay processes by calculating the three-point correlation functions in the framework of the QCD sum rules: $T^a_{c\bar s0}(2900)^{++}\rightarrow D_s^+\pi^+$, $D^+K^+, D_s^{\ast +}\rho^+$, $D_{s1}^+\pi^+$. The first three decay processes are in S-wave decay modes while the last one is in P-wave. Inspired by the accurate prediction on the mass of $T^a_{c\bar s0}(2900)^{++}$ in Ref.~\cite{Chen:2017rhl}, we study these decays by using the interpolating current $J_{5\mu\nu}$ (actually its charge conjugate current with two positive charges) in Eq.~\eqref{currenttensor0+} and considering $T^a_{c\bar s0}(2900)^{++}$ as a $cu\bar d\bar s$ compact tetraquark with $I(J^P)=1(0^+)$. As an isospin partner, the neutral state $T^a_{c\bar s0}(2900)^{0}$ shall have the same mass and decay width with $T^a_{c\bar s0}(2900)^{++}$ in our calculations. The light-cone sum rule is another useful tool to study the strong couplings and hadron properties~\cite{Agaev:2022eyk,Khodjamirian:2011jp,Wang:2020yvi,Li:2020rcg,Khodjamirian:2020mlb}. We shall also consider to investigate the strong decay properties of $T^a_{c\bar s0}(2900)^{++/0}$ by using the light-cone sum rules in the near future. 

This paper is organized as follows. In Sec.~\ref{sec2}, we use QCD sum rules to study the three-point correlation functions for $T^{a++}_{c\bar s0}D_s^+\pi^+$, $T^{a++}_{c\bar s0}D^+K^+$, $T^{a++}_{c\bar s0}D_s^{\ast+}\rho^+$, $T^{a++}_{c\bar s0}D_{s1}^+\pi^+$ vertices. We will study the operator product expansion (OPE) series up to dimension 8 condensates with various tensor structures, and accordingly calculate the coupling constants and partial decay widths of these channels. Finally, we summarize our results on the decay behavior of $T^a_{c\bar s0}(2900)^{++/0}$. 

\section{QCD sum rules and three-point correlation function}\label{sec2}

Over past several decades, the method of QCD sum rules has been proven to be very powerful to study hadron properties~\cite{Shifman:1978bx,Reinders1985,Colangelo2000,Narison:2002woh,Gubler:2018ctz}. In this work, we shall study the three-point correlation function of two-body strong decay process $T^a_{c\bar s0}(2900)^{++}\to A+B$ as 
\begin{align}
\Pi(p, p^{\prime}, q)=&\int
d^4xd^4y\,e^{ip^{\prime} \cdot x}e^{iq \cdot y}\times \nonumber\\
&\langle0|{\mathbb T}[J_A(x)J_B(y)J_{T^a_{c\bar s0}}^\dag(0)]|0\rangle,   \label{three_point_function}
\end{align}
where $J_{T^a_{c\bar s0}}$ is the interpolating current for the $T^a_{c\bar s0}(2900)^{++}$ state while $J_A$ and $J_B$ are the currents for the final states $A$ and $B$, respectively. 
The three point function in Eq.~\eqref{three_point_function} associates the momentum $p$ with $T^a_{c\bar s0}$, $q$ with $B$, and $p'=p-q$ with $A$. 
We shall consider the $T^a_{c\bar s0}(2900)^{++}\to D_s^+\pi^+$, $D^+K^+, D_s^{\ast +}\rho^+$, $D_{s1}^+\pi^+$ decay processes in the following. 

We use the following interpolating current for $T^a_{c\bar s0}(2900)^{++}$ by considering it as a $cu\bar d\bar s$ compact tetraquark with $I(J^P)=1(0^+)$
\begin{align}
J_{T^{a++}_{c\bar s0}}^{\mu\nu}=c^T_aC\gamma^\mu u_b(\bar{d}_{a}\gamma^\nu
C\bar{s}^T_b-\bar{d}_{b}\gamma^\nu C\bar{s}^T_a)\, . \label{currentTcs}
\end{align}
As mentioned above, the hadron mass extracted from this current is $2.88\pm 0.15$ GeV in Ref.~\cite{Chen:2017rhl}, which is in excellent agreement with the observed mass of $T^a_{c\bar s0}(2900)^{++}$~\cite{LHCb:2022sfr,LHCb:2022lzp}. 
The current in Eq.~\eqref{currentTcs} can couple to the $T^a_{c\bar s0}(2900)^{++}$ state via
\begin{eqnarray}
\langle 0|J_{T^{a++}_{c\bar s0}}^{\mu\nu}|T^{a++}_{c\bar s0}(p)\rangle=f_{0T}g^{\mu\nu}\, ,
\label{Tcscoupling}
\end{eqnarray}
in which the coupling constant $f_{0T}$ can be determined from the two-point mass sum rules established in Ref.~\cite{Chen:2017rhl}. Using the same QCD parameters and the same working regions of the Borel mass and continuum threshold as Ref.~\cite{Chen:2017rhl}, we can obtain the value of this coupling constant as 
\begin{align}
f_{0T}=(3.0\pm0.4)\times10^{-3}~\text{GeV}^5  . \label{couplingTcs}
\end{align}
In the following analyses, we shall adopt this value to calculate the decay width of the $T^a_{c\bar s0}(2900)^{++}$ state.

\subsection{Decay mode of \texorpdfstring{$T^a_{c\bar s0}(2900)^{++}\to D_s^+\pi^+$}{}}\label{Sec:Dspi}

In this subsection, we study the decay process $T^a_{c\bar s0}(2900)^{++}\rightarrow D_s^+\pi^+$, the mode in which the $T^a_{c\bar s0}(2900)^{++}$ state was observed. To calculate the three-point correlation function in Eq.~\eqref{three_point_function} for the vertex
$T^a_{c\bar s0}(2900)^{++}D_s^+\pi^+$, we need the interpolating currents for $D_s^+$ and $\pi^+$ mesons
\begin{eqnarray}
J_{D_s^+}&=&\bar s_ai\gamma_5c_a,
\label{eq:Ds}
\\
J_{\pi^+}&=&\bar d_ai\gamma_5 u_a.
\label{eq:pi}
\end{eqnarray}
They can couple to the pseudoscalar $D_s^+$ and $\pi^+$ mesons via the following relations~\cite{Chen:2021erj}
\begin{eqnarray}
    \langle0|J_{D_s^+}|D_s^+(p^{\prime})\rangle&=&\frac{f_{D_s}m^2_{D_s}}{m_c+m_s}\equiv \lambda_{D_s}\, ,
    \\
    \langle0|J_{\pi^+}|\pi^+(q)\rangle&=&\frac{f_{\pi}m^2_{\pi}}{m_u+m_d}\equiv \lambda_{\pi}\, ,\label{pirelation}
\end{eqnarray}
where $f_{D_s}$ and $f_{\pi}$ are coupling constants for $D_s^+$ and $\pi^+$, respectively. From the definition of the three-point correlation function, momentum conservation implies $p=p^{\prime}+q$ in the two-body decay process. Using the above two relations and Eq.~\eqref{Tcscoupling}, we can write down the three-point correlation function Eq.~\eqref{three_point_function} on the phenomenological side
\begin{align}\label{eq:DspiPH}
     &\Pi^{\mu\nu({\rm PH})}_{D_s \pi}(p, p^\prime, q) \nonumber\\ 
    =&\int d^4x d^4y  e^{ip^\prime x} e ^{i q y} \langle 0|{\mathbb T} [J_{D_s^+}(x) J_{\pi^+}(y) J^{\mu\nu\dagger}_{T^{a++}_{c\bar s0}}(0)] |0\rangle \nonumber\\
     =  &\frac{ g^{\mu\nu}f_{0T} \lambda_{D_s}\lambda_{\pi}\langle D_s^+(p^\prime)\pi^+(q)|T^{a++}_{c\bar s0}(p)\rangle}{(p^2-m_{T^a_{c\bar s0}}^2 +i\epsilon) (p^{\prime2}-m_{D_s}^2 +i\epsilon)(q^2-m_\pi^2 +i\epsilon)}
      +\cdots\, , 
\end{align}
in which the ``$\cdots$'' represents the contributions from higher excited states. We insert the intermediate hadronic states $\sum\limits_{X} |X\rangle\langle X|$ into the correlation function in the last step of Eq.~\eqref{eq:DspiPH} to obtain a representation of 
$\Pi^{\mu\nu({\rm PH})}_{D_s \pi}(p, p^\prime, q)$ in terms of hadron parameters that correspond to the currents $J_{D_s^+}$, $J_{\pi^+}$ and $J^{\mu\nu}_{T^{a++}_{c\bar s0}}$. 
 The coupling constant $g_{T^a_{c\bar s0}D_s \pi}$ is defined via the effective Lagrangian~\cite{Hooft2008,AbdelRehim2003}
\begin{eqnarray}
    {\cal L}_{T^a_{c\bar s0}D_s\pi}=-g_{T^a_{c\bar s0}D_s \pi} T_{c\bar s0}^{a++} \partial^\mu\pi^+ \partial_\mu D_s^+ + h.c. \, .
\end{eqnarray}
The transition matrix element $\langle D_s^+(p^\prime)\pi^+(q)|T^{a++}_{c\bar s0}(p)\rangle$ in Eq.~\eqref{eq:DspiPH} can be thus obtained as
\begin{eqnarray}
  \langle D_s^+(p^\prime)\pi^+(q)|T^{a++}_{c\bar s0}(p)\rangle=g_{T^a_{c\bar s0}D_s\pi}p^{\prime}\cdot q,\,  \label{vDspi}
\end{eqnarray}
Then the three-point correlation function is written as
\begin{align}
    &\Pi^{\mu\nu({\rm PH})}_{D_s \pi}(p, p^\prime, q)\nonumber\\
    =&\frac{ g^{\mu\nu}f_{0T} \lambda_{D_s}\lambda_{\pi} g_{T^a_{c\bar s0}D_s\pi}p^{\prime}\cdot q}{(p^2-m_{T^a_{c\bar s0}}^2 +i\epsilon) (p^{\prime2}-m_{D_s}^2 +i\epsilon)(q^2-m_\pi^2 +i\epsilon)}+\cdots\, .\label{eq:DspiPH2}
\end{align}

To calculate the three-point correlation function at the quark-gluon level, we adopt the coordinate-space expression for the light quark propagators and
momentum-space expression for the charm quark propagator
\begin{eqnarray} \nonumber
iS_{ab}^q(x)&=& \frac{i\delta_{ab}}{2\pi^2x^4}\hat{x}
-\frac{i}{32\pi^2}\frac{\lambda^n_{ab}}{2}g_sG_{\mu\nu}^n\frac{1}{x^2}(\sigma^{\mu\nu}\hat{x}+\hat{x}\sigma^{\mu\nu})
\\ &&-\frac{\delta_{ab}}{12}\langle\bar{q}q\rangle - \frac{\delta_{ab}x^2}{192}\langle g_s \bar{q}\sigma Gq\rangle, \label{coordinate_propagator}
\\ \nonumber
iS_{ab}^{c}(p) &=&
\frac{i\delta_{ab}(p\!\!\!\slash+m_c)}{p^2-m_c^2}-\frac{i}{4}g_s\frac{\lambda^n_{ab}}{2}G^n_{\mu\nu}\frac{1}{(p^2-m_c^2)^2} \times
\\ && \{\sigma_{\mu\nu}(p\!\!\!\slash+m_c)+(p\!\!\!\slash+m_c)\sigma_{\mu\nu}\}, \label{momentum_propagator}
\end{eqnarray}
where $m_c$ is the mass of the charm quark. We have neglected the chirally-suppressed contributions from the current light quark masses ($m_q=0$ in the chiral limit) in the OPE calculations because they are numerically insignificant. To establish a sum rule for the coupling constant $g_{T^a_{c\bar s0}D_s\pi}$, we will pick out the $1/q^2$ terms with the structure $g^{\mu\nu}p^{\prime}\cdot q$ in the OPE series and then equate it with the three-point correlation 
function in Eq.~\eqref{eq:DspiPH2} at the hadron level. For this tensor structure, it is found that the quark condensate $\qq$, the quark-gluon mixed condensate $\qGqa$ and the $\GGb\qq$ condensate give contributions to the OPE side at order $1/q^2$ ($q^2$ pole)
\begin{align}
    \rho_{_{D_s\pi}}(s)=&\left((m_c-m_s)^2-s\right)(m_c+m_s)\nonumber\\
    &\times\left(\frac{\qq f(s)}{8\pi^2 s}-\frac{\qGqa }{96\pi^2 s^2 f(s)}\right)\,,\label{spectral_densityDspi}
\end{align}
\begin{equation}
  \begin{split}
    \Pi_{_{D_s\pi}}^{\GGb\qq}(p'^2)=&\frac{\GGb\qq}{288 \pi ^2}\int_{0}^{1}dx\frac{1}{ \Delta(p'^2)^3 (x-1)^2 x^2}\nonumber\\
    &\times \left(2 m_c^3 x^3+2 m_c^2 m_s \left(x^2-x+1\right) x\right. \nonumber\\
    &-\Delta(p'^2)  m_s (x-1)^2 \left(x^2+2\right)\nonumber\\
    &+2 m_c m_s^2 (x-1)^2\nonumber\\
    & -\Delta(p'^2) m_c \left(x^2-2 x+3\right) x^2\bigg) \,,\label{ope_densityDspi_dim7}
  \end{split}
\end{equation}
where
\begin{align}
    f(s)\equiv &\sqrt{\left(1-\frac{(m_c-m_s)^2}{s}\right)\left(1-\frac{(m_c+m_s)^2}{s}\right)}\, ,\\
    \Delta(p'^2)\equiv& \frac{m_c^2}{1-x}-p'^2\, ,
\end{align}
in which $m_s$ is the strange quark mass.

The three-point correlation function in the OPE side can be written as 
\begin{equation}
  \Pi^{\mu\nu({\rm OPE})}_{D_s \pi}=\frac{g^{\mu\nu}p^{\prime}\cdot q}{q^2}\left[\int_{<}^{s_0}ds \frac{\rho^{OPE}(s)}{s-p'^2}+\int_{s_0}^{\infty}ds \frac{\rho^{OPE}(s)}{s-p'^2}\right]\,.\label{eq:DspiOPE1}
\end{equation}
Comparing to Eq.~\eqref{eq:DspiPH2}, it is assumed that the contributions from the higher excited states of ``$\cdots$'' can be compensated by the second term of Eq.~\eqref{eq:DspiOPE1} above the continuum threshold $s_0$, as usual in the QCD sum rule method. Using the soft-pion approximation~\cite{Reinders1985} in Eq.~\eqref{eq:DspiPH2}, one can set up $p^2=p^{\prime 2}=P^2$  ($q\sim 0$) and derive a sum rule for the coupling constant $g_{T^a_{c\bar s0}D_s \pi}$ at the $q^2$ pole after performing the Borel transform ($P^2\to M_B^2$) on both the phenomenological and  OPE sides
\begin{align}
\label{g_Dspi}
g_{T^a_{c\bar s0}D_s \pi}(s_0, M_B^2)
= &\frac{1}{f_{0T} \lambda_{D_s}\lambda_{\pi}}\frac{m^2_{T^a_{c\bar s0}}-m^2_{D_s}}{e^{-m^2_{D_s}/M_B^2}-e^{-m^2_{T^a_{c\bar s0}}/M_B^2}}\nonumber\\
&\times\Bigg(\int_{(m_c+m_s)^2}^{s_0}\rho_{_{D_s\pi}}(s)e^{-s/M_B^2}ds\nonumber\\
& +\Pi_{_{D_s\pi}}^{\GGb\qq}(M_B^2)\Bigg)\, ,
\end{align}
in which $M^2_B$ and $s_0$ are the Borel parameter and continuum threshold, respectively. $\Pi_{_{D_s\pi}}^{\GGb\qq}(M_B^2)$ is the Borel transformation of 
$\Pi_{_{D_s\pi}}^{\GGb\qq}(p'^2)$. 

To perform the QCD sum-rule numerical analysis, the following QCD parameter values of quark masses and various condensates are adopted~\citep{ParticleDataGroup:2022pth,Reinders1985,Narison2012,Kuhn2007}
\begin{equation}\label{condensatepara}
    \begin{split}
        &m_u=2.16^{+0.49}_{-0.26} ~\text{MeV}\, ,\\
        &m_d=4.67^{+0.48}_{-0.17} ~\text{MeV}\, ,\\
        &m_c(\mu=m_c)=\overline{m}_c=1.27\pm 0.02~\text{GeV}\, , \\
        &m_c/m_s=11.76^{+0.05}_{-0.10} \, , \\
        &\qq =-(0.24\pm 0.01)^3~\text{GeV}^3\, ,\\
        &\ssc=(0.8\pm 0.1)\qq\, ,\\
        &\qGqa=M_0^2\qq\, ,\\
        &\sGsa=M_0^2\ssc\, ,\\
        &M_0^2=0.8\pm 0.2~\text{GeV}^2\, ,\\
        &\GGb=0.48\pm 0.14~\text{GeV}^4\, ,
    \end{split}
\end{equation}
where the mass of the charm quark is the ``running'' mass in the $\overline{\text{MS}}$ scheme. We use the renormalization scheme and scale independent $m_c/m_s$ mass ratio from PDG~\cite{ParticleDataGroup:2022pth} to make sure that the strange quark mass $m_s$ is at the same renormalization scale with $m_c$.

\begin{figure}
  \includegraphics[width=8cm,height=5cm]{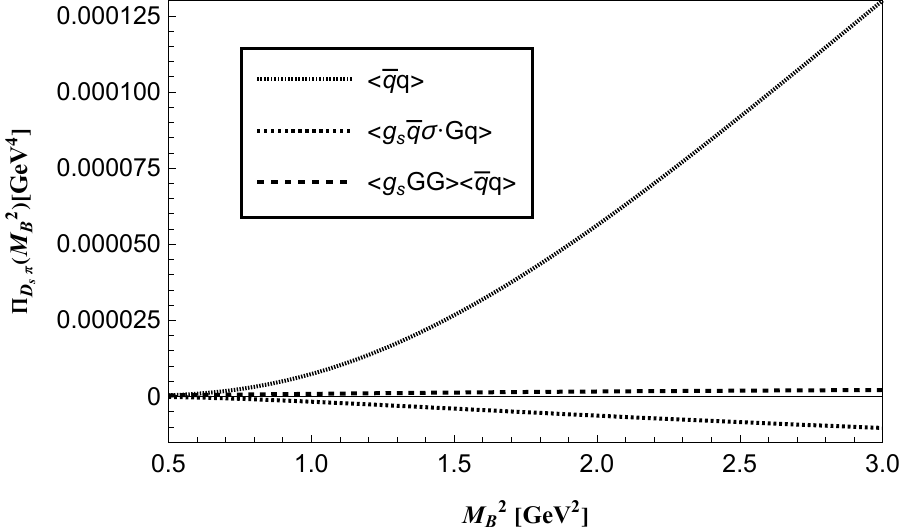}
  \caption{\label{fig7}The contributions of each term in OPE series to the correlation function.}
\end{figure}

In FIG.~\ref{fig7}, we show the contributions of each term in OPE series to the correlation function after doing the Borel transform. It is shown that the dominant contribution is from the quark condensate $\qq$, while the $\GGb\qq$ term is much smaller.

There are also some hadronic parameters~\citep{ParticleDataGroup:2022pth} 
\begin{equation}\label{Dspipara}
    \begin{split}
        m_\pi&=139.57094\pm 0.00018~\text{MeV}, \\
        f_\pi&=130.56~\text{MeV} , \\
        m_{D_s}&=1968.35\pm 0.07~\text{MeV},\\
        f_{D_s}&=246.7~\text{MeV} ,\\
        m_{T^a_{c\bar s0}}&=2908\pm 11\pm 20~\text{MeV},
    \end{split}
\end{equation}
in which we use the experimental value of the mass of $T^a_{c\bar s0}$ state~\cite{LHCb:2022sfr,LHCb:2022lzp}. 

\begin{figure}
    \includegraphics[width=7cm,height=5cm]{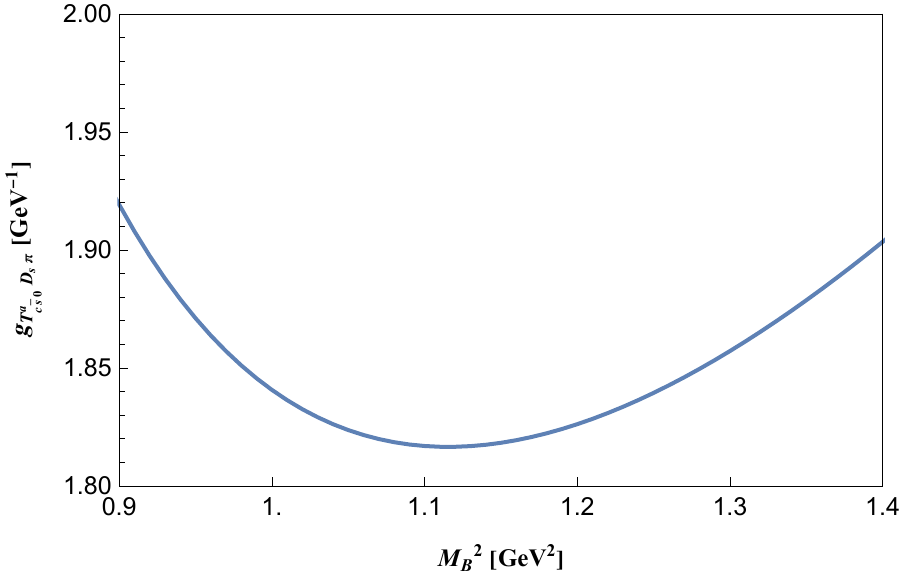}
    \caption{\label{fig1}The dependence of the coupling constant $g_{T^a_{c\bar s0}D_s \pi}$ on the Borel parameter $M_B^2$.}
\end{figure}

In Eq.~\eqref{g_Dspi}, the coupling constant $g_{T^a_{c\bar s0}D_s \pi}$ depends on the continuum threshold $s_0$ and the non-physical parameter $M_B^2$ in the limit $Q^2\to 0$. The continuum threshold $s_0$ as a physical parameter should have the same value as in the mass sum rule in Ref.~\cite{Chen:2017rhl}, i.e., $s_0=10~\text{GeV}^2$. With this value of $s_0$ and the parameters given in Eq.~\eqref{condensatepara} and Eq.~\eqref{Dspipara}, we show the variation of the coupling constant $g_{T^a_{c\bar s0}D_s \pi}$ with the Borel parameter $M_B^2$ in FIG.~\ref{fig1}. We find that the sum rule gives a minimum value of the coupling constant $g_{T^a_{c\bar s0}D_s \pi}$ at $M_B^2\sim 1.12~\text{GeV}^2$, around which it has minimal dependence on the non-physical parameter $M_B^2$. Considering the errors from all parameters, we obtain the coupling constant as
\begin{equation}\label{couplinggDspi}
    g_{T^a_{c\bar s0}D_s \pi}=1.82\pm 0.52~\text{GeV}^{-1}\, .
\end{equation}

The decay width of the two-body strong decay process $T^{a++}_{c\bar s0}\to D_s^+\pi^+$ ($S\to PP$) can be derived from Eq.~\eqref{vDspi} as
\begin{align}
    \Gamma (T^{a++}_{c\bar s0}\rightarrow D_s^+\pi^+)=&\frac{g^2_{T^a_{c\bar s0}D_s \pi}p^\ast (m_{T^a_{c\bar s0}},m_{D_s},m_\pi)}{32 \pi m^2_{T_{c\bar s}}}\times\nonumber\\
    &\left(m^2_{T^a_{c\bar s0}}-m^2_{D_s}-m^2_\pi\right)^2\,, \label{widthformulaDspi}
\end{align}
where the magnitude of outgoing momentum $p^\ast(a,b,c)$ is defined as
\begin{equation}
    p^\ast(a,b,c)\equiv \frac{\sqrt{a^4+b^4+c^4-2a^2b^2-2b^2c^2-2c^2a^2}}{2a}.
\end{equation}

Then we can calculate the partial decay width of the process $T^{a++}_{c\bar s0}\to D_s^+\pi^+$ as
\begin{equation}\label{widthDspi}
    \Gamma (T^{a++}_{c\bar s0}\rightarrow D_s^+\pi^+)=62.9\pm 36.7~\text{MeV}\, ,
\end{equation}
where the error is mainly from the uncertainties of the quark condensate and the quark-gluon mixed condensate.

\subsection{Decay mode of \texorpdfstring{$T^a_{c\bar s0}(2900)^{++}\to D^+K^+$}{}}\label{Sec:DK}
In this subsection, we study the decay process $T^a_{c\bar s0}(2900)^{++}\to D^+K^+$, which is also the two pseudoscalar mesons channel. We use the following pseudoscalar interpolating currents for $D$ and $K$ mesons:
\begin{eqnarray}
    J_{D^+}=\bar{d}_ai\gamma_5c_a\, ,\label{currentD}\\
    J_{K^+}=\bar{s}_ai\gamma_5u_a\,,\label{currentK}
\end{eqnarray}
with the current-meson coupling 
\begin{eqnarray}
    \langle0|J_{D^+}|D^+(p^{\prime})\rangle&=&\frac{f_{D}m^2_{D}}{m_c+m_d}\equiv \lambda_{D}\, ,
    \\
    \langle0|J_{K^+}|K^+(q)\rangle&=&\frac{f_{K}m^2_{K}}{m_u+m_s}\equiv \lambda_{K}\, ,
\end{eqnarray}
where $f_{D}$ and $f_{K}$ are the decay constants for $D$ and $K$ mesons, respectively. Then the three-point function at the hadron level can be written as 
\begin{align}
    &\Pi^{\mu\nu({\rm PH})}_{D K}(p, p^\prime, q)\nonumber\\
    =&\int d^4x d^4y  e^{ip^\prime x} e ^{i q y} \langle 0|{\mathbb T} [J_{D^+}(x) J_{K^+}(y) J^{\mu\nu\dagger}_{T^{a++}_{c\bar s0}}(0) ]|0\rangle \nonumber\\
    =  &\frac{g^{\mu\nu}f_{0T} \lambda_{D}\lambda_{K}\langle D^+(p^\prime)K^+(q)|T^{a++}_{c\bar s0}(p)\rangle}{(p^2-m_{T^a_{c\bar s0}}^2 +i\epsilon) (p^{\prime2}-m_{D}^2 +i\epsilon)(q^2-m_K^2 +i\epsilon)}
  +\cdots\nonumber\\
    =&\frac{ g^{\mu\nu}f_{0T} \lambda_{D}\lambda_{K} g_{T^a_{c\bar s0}DK}p^{\prime}\cdot q}{(p^2-m_{T^a_{c\bar s0}}^2 +i\epsilon) (p^{\prime2}-m_{D}^2 +i\epsilon)(q^2-m_K^2 +i\epsilon)}+\cdots\, ,\label{eq:DKPH2}
\end{align}
where the coupling constant $g_{T^a_{c\bar s0}DK}$ is defined in the same way as $g_{T^a_{c\bar s0}D_s\pi}$~\cite{Hooft2008,AbdelRehim2003}
\begin{equation}
    {\cal L}_{T^a_{c\bar s0}DK}=-g_{T^a_{c\bar s0}DK} T_{c\bar s0}^{a++} \partial^\mu K^+ \partial_\mu D^+ + h.c. \, . \\
\end{equation}

The calculation of the three-point correlation function at the quark-gluon level for this decay mode is similar to the $D_s\pi$ mode. However, we cannot simply ignore the mass of $K$ meson in this situation as we did for $\pi$ meson. To apply sum rules appropriately, we still extract the terms proportional to $1/q^2$ with the structure $g^{\mu\nu}p^{\prime}\cdot q$ in the OPE series, implying that the coupling constant $g_{T^a_{c\bar s0}DK}$ as a function of $Q^2(=-q^2)$ is calculated at $Q^2$ far away from the on-shell mass $-m_K^2$. Then we will extrapolate the coupling constant $g_{T^a_{c\bar s0}DK}$ at $Q^2=-m_K^2$ from the QCD sum rule working region~\cite{Chen2015,Dias2013}.

We obtain the following OPE for this channel
\begin{align}
    \rho_{_{DK}}(s)=&-\frac{m_c(m_c^2-s)^2(\qq+\ssc)}{16\pi^2s^2}\nonumber\\
    &+\frac{m_c^3\left(\qGqa-\sGsa\right)}{192\pi^2s^2}\nonumber\\
    &+\frac{m_s\qGqa \delta(s-m_c^2)}{192\pi^2}+\frac{m_c\sGsa}{96\pi^2s}\nonumber\\
    &+\frac{\GGb m_cm_s(2m_c^2-7s)}{4608\pi^2s^2}\, , \label{spectral_densityDK}
\end{align}
\begin{align}
    \Pi_{_{DK}}^{\GGb\qq}(p'^2)=&\frac{\GGb\qq m_s}{1152\pi^2(p'^2-m_c^2)^2}+\nonumber\\
    &\frac{\GGb (\qq+\ssc) m_c}{576 \pi ^2}\int_{0}^{1}\frac{dx}{ \Delta(p'^2) ^3 (x-1)^2}\nonumber\\
    &\times \left(2 m_c^2 x-\Delta(p'^2)  \left(x^2-2 x+3\right)\right)\, , \label{ope_densityDK_dim7}
\end{align}
which contain the quark condensates, quark-gluon mixed condensates and gluon condensates. 
Then we can establish the sum rule for $g_{T^a_{c\bar s0}DK}$ by assuming $p^2=p^{\prime 2}=P^2$ and performing the Borel transform on both the phenomenological and OPE sides 
\begin{align}
    g_{T^a_{c\bar s0}DK}(s_0,M_B^2,Q^2)=&\frac{1}{f_{0T}\lambda_D\lambda_K}\frac{m^2_{T^a_{c\bar{s}0}}-m_D^2}{e^{-m^2_{D}/M_B^2}-e^{-m^2_{T^a_{c\bar s0}}/M_B^2}}\nonumber\\
    &\times\left(\frac{Q^2+m_K^2}{Q^2}\right)\Bigg(\int_{m_c^2}^{s_0}\rho_{_{DK}}(s)e^{-s/M_B^2}ds\nonumber\\
    &+\Pi_{_{DK}}^{\GGb\qq}(M_B^2)\Bigg)\, .
\end{align}

The following parameters for the $D$ and $K$ mesons are adopted to perform numerical analysis~\cite{ParticleDataGroup:2022pth}
\begin{equation}\label{DKpara}
    \begin{split}
        m_K=493.677\pm 0.016~\text{MeV},\quad &f_K=155.7~\text{MeV},\\
        m_{D}=1869.66\pm 0.05~\text{MeV},\quad &f_{D}=208.9~\text{MeV}.
    \end{split}
\end{equation}
\begin{figure}
    \includegraphics[width=7cm,height=5cm]{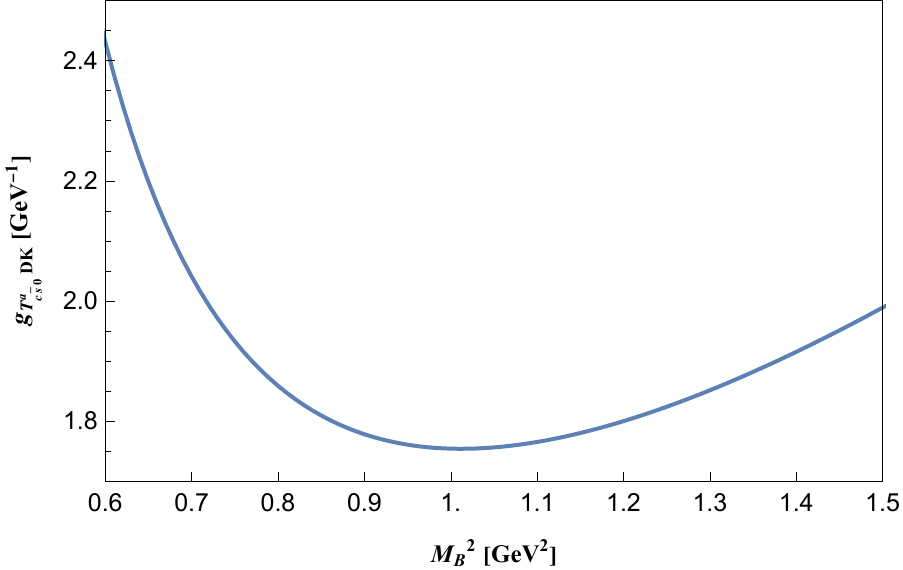}
    \caption{\label{fig2}Variation of the coupling constant $g_{T^a_{c\bar s0}DK}(Q^2)$ at $Q^2=3.5~\text{GeV}^2$ with the Borel parameter $M_B^2$.}
\end{figure}

\begin{figure}
    \includegraphics[width=7cm,height=5cm]{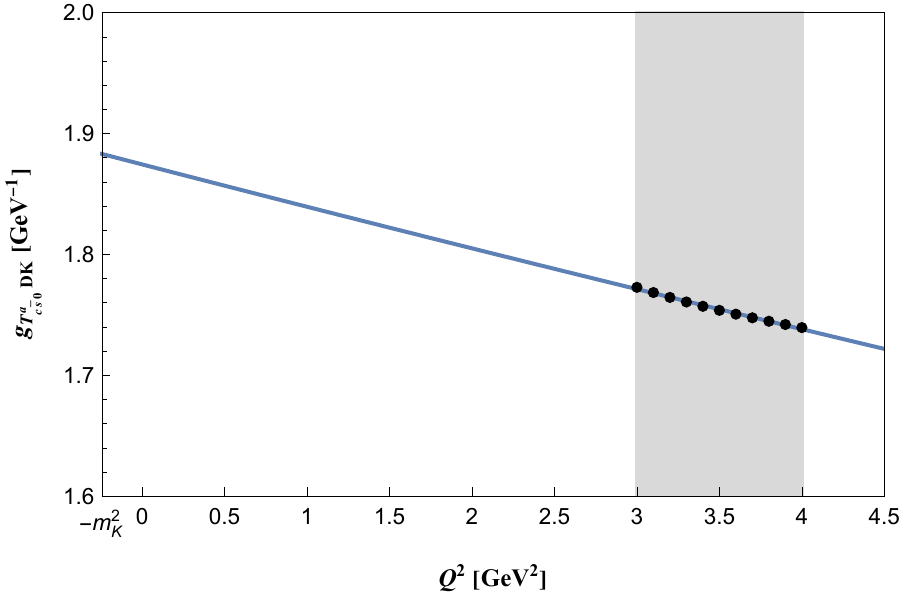}
    \caption{\label{fig3}Variation of the coupling constant $g_{T^a_{c\bar s0}DK}$ with $Q^2$. The dotted points represent the QCD sum rule results for $g_{T^a_{c\bar s0}DK}$ with $s_0=10~\text{GeV}^2$ and $M_B^2=1.01~\text{GeV}^2$. The solid line is the fit of the QCD sum rule result through model Eq.~\eqref{DKfit} and the extrapolation to the physical pole $Q^2=-m_K^2$.}
\end{figure}

In FIG.~\ref{fig2}, we show the variation of the coupling constant $g_{T^a_{c\bar s0}DK}(Q^2)$ with the Borel parameter $M_B^2$ at the Euclidean momentum point $Q^2=m_D^2\approx 3.5~\text{GeV}^2$ and the continuum threshold $s_0=10~\text{GeV}^2$. Such a momentum point is chosen far away from  $m_K^2$ so that it can be safely ignored and the OPE series is valid in this region. We can see that the minimum value of $g_{T^a_{c\bar s0}DK}$ appears around $M_B^2=1.01~\text{GeV}^2$. To obtain the value of $g_{T^a_{c\bar s0}DK}$ at $Q^2=-m_K^2$, we extrapolate the coupling constant $g_{T^a_{c\bar s0}DK}$ from the valid QCD sum rule working region to the physical pole $Q^2=-m_K^2$ by using the exponential model~\cite{Chen2015,Dias2013}
\begin{equation}\label{DKfit}
    g_{T^a_{c\bar s0}DK}=g_1e^{-g_2Q^2}\,,
\end{equation}
where the parameters $g_1$ and $g_2$ can be determined by fitting the QCD sum rule result for $g_{T^a_{c\bar s0}DK}$. In FIG.~\ref{fig3}, we fit the QCD sum rule result for $s_0=10~\text{GeV}^2$ and $M_B^2=1.01~\text{GeV}^2$ well described by the model Eq.~\eqref{DKfit} when $g_1=1.87~\text{GeV}^{-1}$ and $g_2=0.02~\text{GeV}^{-2}$. Then we can obtain the value of the coupling constant $g_{T^a_{c\bar s0}DK}$ at the physical pole $Q^2=-m_K^2$
\begin{equation}\label{DKcoupling}
    g_{T^a_{c\bar s0}DK}(Q^2=-m_K^2)=1.88\pm 0.55~\text{GeV}^{-1}\, .
\end{equation}
As the $S\to PP$ decay process, the decay width for $T^a_{c\bar s0}(2900)^{++}\to D^+K^+$ can be evaluated by  Eq.~\eqref{widthformulaDspi}. Using the parameter values mentioned above, we obtain the partial decay width of this process as 
\begin{equation}\label{widthDK}
    \Gamma(T_{c\bar s0}^{a++}\rightarrow D^+K^+)=69.2\pm 40.8~\text{MeV}\, ,
\end{equation}
where the error is mainly from the uncertainties of the quark condensates $\qq$ and $\ssc$.
\subsection{Decay mode of \texorpdfstring{$T^a_{c\bar s0}(2900)^{++}\to D_s^{\ast +} \rho^+$}{}}\label{Sec:Dsstarrho}
In this subsection, we study the decay mode $T^a_{c\bar s0}(2900)^{++}\to D_s^{\ast +} \rho^+$ as an $S\to VV$ process. We use the vector interpolating currents for the $D_s^{\ast +}$ and $\rho^+$ mesons  
\begin{eqnarray}
    J_{D_s^{\ast +}}^\alpha&=&\bar{s}_a\gamma^\alpha c_a\,,\label{currentDsstar}\\
    J_{\rho^+}^\beta&=&\bar d_a\gamma^\beta u_a\, . \label{currentrho}
\end{eqnarray}
The corresponding current-meson couplings are 
\begin{eqnarray}
    \langle0|J_{D_s^{\ast +}}^\alpha|D_s^{\ast +}(p^{\prime})\rangle&=&f_{D^{\ast }_s}m_{D^{\ast }_s}\epsilon^\alpha(p^\prime)=\lambda_{D_s^{\ast}}\epsilon^\alpha(p^\prime)\, ,
    \\
    \langle0|J_{\rho^+}^\beta|\rho^+(q)\rangle&=&f_{\rho}m_{\rho}\epsilon^\beta(q)=\lambda_{\rho}\epsilon^\beta(q)\, ,
\end{eqnarray}
where $f_{D^{\ast }_s}$ and $\epsilon^\alpha(p^\prime)$ are the decay constant and polarization vector for $D^{\ast }_s$ respectively, while $f_{\rho}$ and $\epsilon^\beta(q)$ are for the $\rho$ meson.

To match the structure of the $SVV$ three-point correlation function, one can use the following $T^{a++}_{c\bar s0}D^{\ast +}_s\rho^+$ interaction vertex~\cite{Moussallam1994,Dai2019}
\begin{equation}
    \langle D_s^{*+}(p')\rho^+(q)\vert T^{a++}_{c\bar s 0}(p)\rangle=-g_{T^a_{c\bar s 0}D_s^\ast\rho} \epsilon^{\ast}(p')\cdot\epsilon^{\ast}(q)\,,\label{vDsstarrho}
\end{equation}
in which and the coupling constant $g_{T^a_{c\bar s 0}D_s^\ast\rho}$ is defined through
\begin{equation}
    {\cal L}_{T^a_{c\bar s0}D_s^\ast\rho}=-g_{T^a_{c\bar s 0}D_s^\ast\rho} T_{c\bar s0}^{a++} D_{s}^{\ast +\mu}\rho^{+}_{\mu}+ h.c. \, . \\
\end{equation}

Then the three-point correlation function at the hadron level can be written as 
\begin{align}
    &\Pi^{\mu\nu, \alpha\beta({\rm PH})}_{D_{s}^{\ast} \rho}(p, p^\prime, q)\nonumber\\
    =&\int d^4x d^4y e^{ip^\prime x} e ^{i q y} \langle 0|{\mathbb T} [J_{D_{s}^{\ast+}}^\alpha(x) J_{\rho^+}^\beta(y) J^{\mu\nu\dagger}_{T^{a++}_{c\bar s0}}(0) ]|0\rangle \nonumber\\
    =  &\frac{ g^{\mu\nu}f_{0T} \lambda_{D_{s}^{\ast}}\lambda_{\rho}\epsilon^\alpha(p^\prime)\epsilon^\beta(q)\langle D_s^{\ast^+}(p')\rho^+(q)\vert T^{a++}_{c\bar s 0}(p)\rangle}{(p^2-m_{T^a_{c\bar s0}}^2 +i\epsilon) (p^{\prime2}-m_{D_{s}^{\ast}}^2 +i\epsilon)(q^2-m_\rho^2 +i\epsilon)}
  +\cdots\nonumber\\
    =&\frac{ g^{\mu\nu}f_{0T} \lambda_{D_{s}^{\ast}}\lambda_{\rho} g_{T^a_{c\bar s 0}D_s^\ast\rho}\left(\frac{p^{\prime\alpha}p^{\prime\beta}}{p^{\prime2}}+\frac{q^{\alpha}q^{\beta}}{q^{2}}-\frac{p^{\prime\alpha}q^\beta p^\prime\cdot q}{p^{\prime2}q^2}-g^{\alpha\beta}\right)}{(p^2-m_{T^a_{c\bar s0}}^2 +i\epsilon) (p^{\prime2}-m_{D_{s}^{\ast}}^2 +i\epsilon)(q^2-m_\rho^2 +i\epsilon)}+\cdots\, .\label{eq:DsstarrhoPH2}
\end{align}
There are four different tensor structures $p^{\prime\alpha}p^{\prime\beta}$, $q^{\alpha}q^{\beta}$, $p^{\prime\alpha}q^\beta$ and $g^{\alpha\beta}$ in Eq.~\eqref{eq:DsstarrhoPH2}. In our calculations on the OPE side, we shall use the terms proportional to $1/q^2$ with tensor structure $g^{\alpha\beta}$ to match those on the phenomenological side. Then we find that only the gluon condensate $\GGb$ contributes to the spectral density
\begin{equation}
    \rho_{_{D_{s}^{\ast} \rho}}(s)=-\frac{\GGb\left((m_c-m_s)^2-s\right)\left((m_c+m_s)^2+2s\right)f(s)}{2304\pi^2s}\, .
\end{equation}
The QCD sum rules for the coupling $g_{T^a_{c\bar s 0}D_s^\ast\rho}$ can be established after performing Borel transform
\begin{align}
    g_{T^a_{c\bar s 0}D_s^\ast\rho}(s_0,M_B^2,Q^2)=&\frac{1}{f_{0T}\lambda_{D_s^\ast}\lambda_\rho}\frac{m^2_{T^a_{c\bar{s}0}}-m_{D_s^\ast}^2}{e^{-m^2_{D_s^\ast}/M_B^2}-e^{-m^2_{T^a_{c\bar s0}}/M_B^2}}\nonumber\\
    &\times\left(\frac{Q^2+m_\rho^2}{Q^2}\right)\int_{(m_c+m_s)^2}^{s_0}\rho_{_{D_{s}^{\ast} \rho}}(s)e^{-s/M_B^2}ds.
\end{align}

\begin{figure}
    \includegraphics[width=7cm,height=5cm]{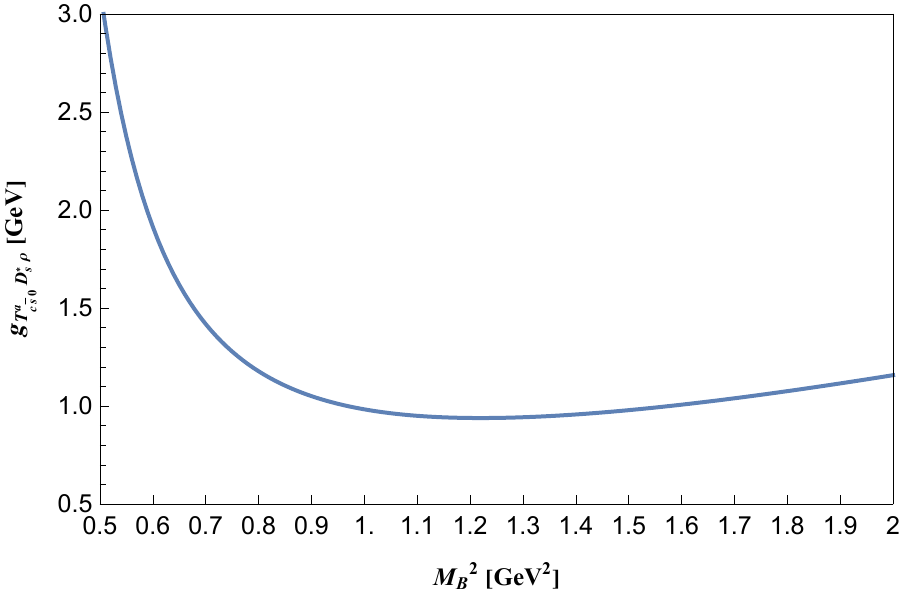}
    \caption{\label{fig4}Variation of the coupling constant $g_{T^a_{c\bar s 0}D_s^\ast\rho}(Q^2)$ at $Q^2=5~\text{GeV}^2$ with the Borel parameter $M_B^2$.}
\end{figure}

\begin{figure}
    \includegraphics[width=7cm,height=5cm]{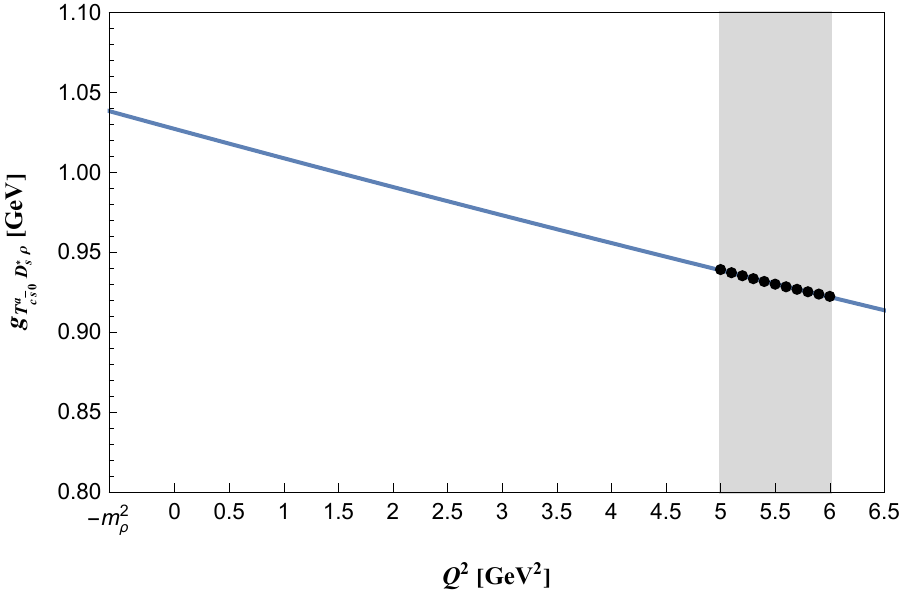}
    \caption{\label{fig5}Variation of the coupling constant $g_{T^a_{c\bar s 0}D_s^\ast\rho}$ with $Q^2$. The dotted points represent the QCD sum rule results for $g_{T^a_{c\bar s 0}D_s^\ast\rho}$ with $s_0=10~\text{GeV}^2$ and $M_B^2=1.22~\text{GeV}^2$. The solid line is the fit of the QCD sum rule result through model Eq.~\eqref{DKfit} and the extrapolation to the physical pole $Q^2=-m_\rho^2$.}
\end{figure}

The hadron parameters for $D_s^\ast$ and $\rho$ are~\cite{ParticleDataGroup:2022pth,Dias2013,Lucha2014}
\begin{equation}\label{Dsstarrhopara}
    \begin{split}
        m_\rho=775.26\pm 0.23~\text{MeV},\quad &f_\rho=157~\text{MeV},\\
        m_{D_s^\ast}=2112.2\pm 0.4~\text{MeV},\quad &f_{D_s^\ast}=305.5~\text{MeV}.
    \end{split}
\end{equation}
In FIG.~\ref{fig4}, we show the variation of the coupling constant $g_{T^a_{c\bar s 0}D_s^\ast\rho}(Q^2)$ with the Borel parameter $M_B^2$ at $Q^2=5~\text{GeV}^2$, which is far away from $m_\rho^2$. The minimum value of $g_{T^a_{c\bar s 0}D_s^\ast\rho}(Q^2)$ appears around $M_B^2=1.22~\text{GeV}^2$. To extrapolate $g_{T^a_{c\bar s 0}D_s^\ast\rho}$ to the physical pole $Q^2=-m_\rho^2$, we fit the QCD sum rule result with $s_0=10~\text{GeV}^2$ and $M_B^2=1.22~\text{GeV}^2$ by using the exponential model Eq.~\eqref{DKfit} with $g_1=1.02~\text{GeV}$ and $g_2=0.02~\text{GeV}^{-2}$, as shown in FIG.~\ref{fig5}. The coupling constant at the physical pole is obtained as 
\begin{equation}\label{Dsstarrhocoupling}
    g_{T^a_{c\bar s 0}D_s^\ast\rho}(Q^2=-m_\rho^2)=1.04\pm 0.48~\text{GeV}\, .
\end{equation} 
The decay width of $T^a_{c\bar s0}(2900)^{++}\to D_s^{\ast +} \rho^+$ as an $S\to VV$ process can be derived from Eq.~\eqref{vDsstarrho} 
\begin{align}\label{widthformulaDsstarrho}
    \Gamma (T_{c\bar s0}^{a++}\rightarrow D_s^{\ast +} \rho^+)= &\frac{p^\ast (m_{T^a_{c\bar s0}},m_{D_s^{\ast }},m_\rho)}{8\pi}g^2_{T^a_{c\bar s 0}D_s^\ast\rho}\nonumber\\
    &\times\left(\frac{3}{m_{T^a_{c\bar s0}}^2}+\frac{p^\ast (m_{T^a_{c\bar s0}},m_{D_s^{\ast }},m_\rho)^2}{m_{D_s^{\ast }}^2m_\rho^2}\right)\,.
\end{align}
Then the partial decay width of this decay mode can be obtained as 
\begin{equation}\label{widthDsstarrho}
    \Gamma (T_{c\bar s0}^{a++}\rightarrow D_s^{\ast +} \rho^+)=2.4\pm 2.3~\text{MeV}\, ,
\end{equation}
where the error is mainly from the uncertainty of the gluon condensate. This partial decay width is much smaller than those for the $D_s^+\pi^+$ and $D^+K^+$ decay modes due to the suppressed phase space.

\subsection{Decay mode of \texorpdfstring{$T^a_{c\bar s0}(2900)^{++}\to D_{s1}^{+} \pi^+$}{}}\label{Sec:Ds1pi}
In this subsection, we study the P-wave decay mode $T^a_{c\bar s0}(2900)^{++}\to D_{s1}^{+} \pi^+$ ($S\to AP$). The axial-vector interpolating current for $D_{s1}^+$ meson is 
\begin{equation}\label{eq:Ds1}
    J_{D_{s1}^+}^\alpha=\bar{s}\gamma^\alpha\gamma_5c\,,
\end{equation}
with the current-meson coupling 
\begin{equation}
    \langle0|J_{D_{s1}^{+}}^\alpha|D_{s1}^{+}(p^{\prime})\rangle=f_{D_{s1}}m_{D_{s1}}\epsilon^\alpha(p^\prime)=\lambda_{D_{s1}}\epsilon^\alpha(p^\prime)\, .
\end{equation}

The effective Lagrangian of $T_{c\bar s0}^{a++}D_{s1}^+\pi^+$ vertex is constructed as 
\begin{equation}
    {\cal L}_{T^a_{c\bar s0}D_{s1}\pi}=g_{T^a_{c\bar s0}D_{s1} \pi} T_{c\bar s0}^{a++} D_{s1}^{+\mu}\partial_\mu\pi^+ + h.c. \,.
\end{equation}
One can immediately read out the transition matrix element 
\begin{equation}\label{vDs1pi}
    \langle D_{s1}^+(p^\prime)\pi^+(q)|T_{c\bar s0}^{a++}(p)\rangle=i g_{T^a_{c\bar s0}D_{s1} \pi} q\cdot \epsilon^\ast(p^\prime)\, ,
\end{equation}
where $g_{T^a_{c\bar s0}D_{s1} \pi}$ is the coupling constant of the $T_{c\bar s0}^{a++}D_{s1}^+\pi^+$ vertex, and $\epsilon^\ast(p^\prime)$ is the polarization vector of $D_{s1}^+$ meson.
Using these definitions, the three-point correlation function for $T^a_{c\bar s0}(2900)^{++}\to D_{s1}^{+} \pi^+$ process at the hadron level can be written as 
\begin{align}
    &\Pi^{\mu\nu, \alpha({\rm PH})}_{D_{s1} \pi}(p, p^\prime, q)\nonumber\\
    =&\int d^4x d^4y e^{ip^\prime x} e ^{i q y} \langle 0|{\mathbb T} [J_{D_{s1}^+}^\alpha(x) J_{\pi^+}(y) J^{\mu\nu\dagger}_{T^{a++}_{c\bar s0}}(0) ]|0\rangle \nonumber\\
    =  &\frac{ ig^{\mu\nu}f_{0T} \lambda_{D_{s1}}\lambda_{\pi}\epsilon^\alpha(p^\prime)\langle D_{s1}^{+}(p')\pi^+(q)\vert T^{a++}_{c\bar s 0}(p)\rangle}{(p^2-m_{T^a_{c\bar s0}}^2 +i\epsilon) (p^{\prime2}-m_{D_{s1}}^2 +i\epsilon)(q^2-m_\pi^2 +i\epsilon)}
  +\cdots\nonumber\\
    =&\frac{ig^{\mu\nu}f_{0T} \lambda_{D_{s1}}\lambda_{\pi} g_{T^a_{c\bar s0}D_{s1} \pi}\left(-q^\alpha+\frac{p^\prime\cdot qp^{\prime\alpha}}{p^{\prime2}}\right)}{(p^2-m_{T^a_{c\bar s0}}^2 +i\epsilon) (p^{\prime2}-m_{D_{s1}}^2 +i\epsilon)(q^2-m_\pi^2 +i\epsilon)}+\cdots\, .\label{eq:Ds1piPH2}
\end{align}
There are two different tensor structures $q^\alpha$ and $p^{\prime\alpha}$ in the three-point correlation function. 
To study the $S\to AP$ decay mode, we use the $q^\alpha$ structure in the three-point correlation function on both the phenomenological and OPE sides in the $m_\pi\to0$ limit. 
At the quark-gluon level, we obtain the terms proportional to $1/q^2$ in the OPE series as  
\begin{align}
    \rho_{_{D_{s1}\pi}}(s)=&\frac{\qq\left((m_c+m_s)^2-s\right)\left((m_c-m_s)^2+2s\right)f(s)}{24\pi^2s}\nonumber\\
    &-\frac{3\qGqa(m_c-m_s)^2\left((m_c+m_s)^2-s\right)}{288\pi^2s^2f(s)}\, , \label{spectral_densityDs1pi}
\end{align}
\begin{align}
  \Pi_{_{D_{s1}\pi}}^{\GGb\qq}(p'^2)=&\frac{\GGb \qq }{288 \pi ^2}\int_{0}^{1}\frac{dx}{\Delta(p'^2) ^3 (x-1)^3 x^2}\nonumber\\
  &\times\bigg(-2 m_c^4 x^3-2 m_c^3 m_s (x-1) x+\nonumber\\
  &m_c^2 (x-1) \left(2 m_s^2 (x-1)^2+\Delta(p'^2)  x^2 (1-2 x)\right)\nonumber\\
  &+2 \Delta(p'^2)  m_c m_s \left(2 x^3-4 x^2+3 x-1\right)\nonumber\\
  &+\Delta(p'^2)  (x-1)^3 \left(m_s^2 (x-1)+2 \Delta(p'^2) x^2\right)\bigg)\, ,\label{ope_densityDs1pi_dim7}
\end{align}
in which the quark condensate, the quark-gluon mixed condensate and the gluon condensate are included. 

We establish the sum rules for the coupling constant $g_{T^a_{c\bar s0}D_{s1} \pi}$ at the $q^2$ pole as
\begin{align}
    \label{g_Ds1pi}
    &g_{T^a_{c\bar s0}D_s \pi}(s_0, M_B^2)
    = \nonumber\\ &\frac{1}{f_{0T} \lambda_{D_{s1}}\lambda_{\pi}}
    \frac{m^2_{T^a_{c\bar s0}}-m^2_{D_{s1}}}{e^{-m^2_{D_{s1}}/M_B^2}-e^{-m^2_{T^a_{c\bar s0}}/M_B^2}}\Bigg(\int_{(m_c+m_s)^2}^{s_0}\rho_{_{D_{s1}\pi}}(s)e^{-s/M_B^2}ds\nonumber\\
    &+\Pi_{_{D_{s1}\pi}}^{\GGb\qq}(M_B^2) \Bigg)\, .
\end{align}
\begin{figure}
    \includegraphics[width=7cm,height=5cm]{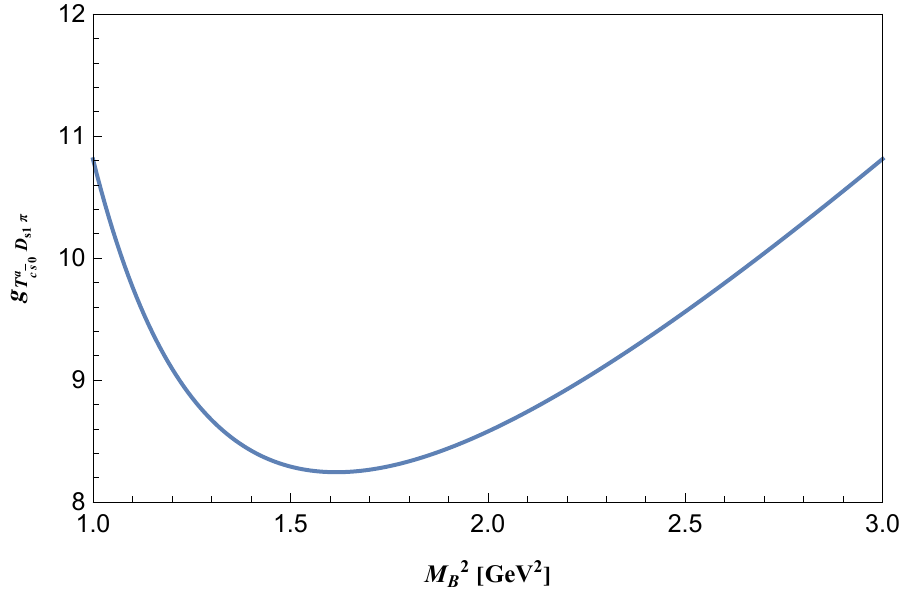}
    \caption{\label{fig6}The dependence of the coupling constant $g_{T^a_{c\bar s0}D_{s1} \pi}$ on the Borel parameter $M_B^2$.}
\end{figure}

To perform numerical analysis, the hadron parameters for the $D_{s1}$ meson are~\cite{ParticleDataGroup:2022pth,Wang2015}
\begin{equation}\label{Ds1para}
    \begin{split}
        m_{D_{s1}}=2459.5\pm 0.6~\text{MeV},\quad &f_{D_{s1}}=345~\text{MeV}\, ,
    \end{split}
\end{equation}
in which the coupling constant $f_{D_{s1}}$ is the theoretical predicted value by QCD sum rule method~\cite{Wang2015}. 

In FIG.~\ref{fig6}, we show the variation of the coupling constant $g_{T^a_{c\bar s0}D_{s1} \pi}$ with $M_B^2$ for $s_0=10~\text{GeV}^2$, in which  the minimal dependence of $g_{T^a_{c\bar s0}D_{s1} \pi}$ on $M_B^2$ appears around $M_B^2\sim 1.62~\text{GeV}^2$. 
The coupling constant at this point is 
\begin{equation}\label{couplingDs1pi}
    g_{T^a_{c\bar s0}D_{s1} \pi}=8.24\pm 2.23\, .
\end{equation}

The decay width of $T^a_{c\bar s0}(2900)^{++}\to D_{s1}^{+} \pi^+$ as an $S\to AP$ process can be derived from Eq.~\eqref{vDs1pi} as
\begin{equation}\label{widthformulaDs1pi}
    \Gamma(T^{a++}_{c\bar s0}\to D_{s1}^{+} \pi^+)=\frac{p^\ast (m_{T^a_{c\bar s0}},m_{D_{s1}},m_\pi)^3}{8\pi m^2_{D_{s1}}}g^2_{T^a_{c\bar s0}D_{s1} \pi}\, ,
\end{equation}
from which the numerical value of this partial decay width is 
\begin{equation}\label{widthDs1pi}
    \Gamma(T^{a++}_{c\bar s0}\to D_{s1}^{+} \pi^+)=27.2\pm15.0~\text{MeV}\, ,
\end{equation}
where the error is mainly from the uncertainty of the quark condensate. This partial decay width is surprisingly large because the phase space in this channel is sizable, although it is in P-wave.

\section{Conclusions and discussions}
In this work, we have investigated the recent observed resonance $T^a_{c\bar s0}(2900)^{++}$ ($T^a_{c\bar s0}(2900)^{0}$) as a fully open-flavor $cu\bar{d}\bar{s}$ ($cd\bar{u}\bar{s}$) tetraquark state with $I(J^P)=1(0^+)$. 
In Ref.~\cite{Chen:2017rhl}, we have successfully predicted the existence of such open-flavor tetraquark 
states by studying their mass spectra and suggested searching for them in the $D_s\pi$ final states, which are exactly the observed channels of $T^a_{c\bar s0}(2900)^{++}$ and $T^a_{c\bar s0}(2900)^{0}$. To further understand the  
nature of $T^a_{c\bar s0}(2900)^{++}$, we have studied its two-body strong decay processes $T^a_{c\bar s0}(2900)^{++}\rightarrow D_s^+\pi^+$, $D^+K^+, D_s^{\ast +}\rho^+$, $D_{s1}^+\pi^+$ in the compact tetraquark picture using the method of QCD sum rules.

We calculated the three-point correlation functions up to dimension eight condensates for the $T^{a++}_{c\bar s0}D_s^+\pi^+$, $T^{a++}_{c\bar s0}D^+K^+$, $T^{a++}_{c\bar s0}D_s^{\ast+}\rho^+$, $T^{a++}_{c\bar s0}D_{s1}^+\pi^+$ vertices. By extracting the $1/Q^2$ terms in the OPE series, we can establish QCD sum rules for the 
coupling constant and then extrapolate its value to the physical pole. After the numerical analyses, we obtain the partial decay widths of the four decay channels as
\begin{equation}\label{decaywidth}
    \begin{split}
        &\Gamma (T^{a++}_{c\bar s0}\rightarrow D_s^+\pi^+)=62.9\pm 36.7~\text{MeV}\, ,\\
        &\Gamma(T_{c\bar s0}^{a++}\rightarrow D^+K^+)=69.2\pm 40.8~\text{MeV}\,,\\
        &\Gamma (T_{c\bar s0}^{a++}\rightarrow D_s^{\ast +} \rho^+)=2.4\pm 2.3~\text{MeV}\, ,\\
        &\Gamma(T^{a++}_{c\bar s0}\to D_{s1}^{+} \pi^+)=27.2\pm 15.0~\text{MeV}\, .
    \end{split}
\end{equation}
The resulting full width of $T^a_{c\bar s0}(2900)^{++}$ is 
\begin{equation}\label{Tcswidth}
    \Gamma_{T^{a++}_{c\bar s0}}=161.7\pm 94.8~\text{MeV}\,,
\end{equation}
which is in good accord with the experimental value~\cite{LHCb:2022sfr,LHCb:2022lzp}. This result supports the compact tetraquark explanation of $T^a_{c\bar s0}(2900)^{++}$ and $T^a_{c\bar s0}(2900)^{0}$, as predicted in Ref.~\cite{Chen:2017rhl}. 

It is useful to give the relative branching ratios from Eq.~\eqref{decaywidth}
\begin{align}\label{decayratio}
\nonumber &\Gamma (T^{a++}_{c\bar s0}\rightarrow D_s^+\pi^+):
\Gamma(T_{c\bar s0}^{a++}\rightarrow D^+K^+):
\Gamma (T_{c\bar s0}^{a++}\rightarrow D_s^{\ast +} \rho^+): \\
&
\Gamma(T^{a++}_{c\bar s0}\to D_{s1}^{+} \pi^+)\approx1.00:1.10:0.04:0.43\, ,
\end{align}
where only the central values of these partial decay widths are adopted. These relative branching ratios can vary in a broad range considering the uncertainties of the partial decay widths in Eq.~\eqref{decaywidth}. Our results show that the main decay modes of $T^a_{c\bar s0}(2900)^{++/0}$ state are $D_s\pi$ and $DK$ channels. However, the $P$-wave decay mode $D_{s1}\pi$ is also comparable and important by including the uncertainties. The investigations of nonrelativistic potential quark model also show that the main decay modes are the same with our results with similar relative branching ratios $\Gamma (T^{a++}_{c\bar s0}\rightarrow D_s\pi):
\Gamma(T^{a++}_{c\bar s0}\rightarrow DK):\Gamma (T^{a++}_{c\bar s0}\rightarrow D_s^{\ast} \rho)=1.00:1.41:0.01$~\cite{Liu:2022hbk}, although their total decay width is much smaller.
By considering $T^a_{c\bar s0}(2900)^{++/0}$ as a $D^{\ast}K^{\ast}$ molecule, the decay behavior is also very different because the partial width of the $DK$ channel is much larger than those in other channels~\cite{Yue:2022mnf}. 

To conclude this paper, we suggest confirming these exotic $T^a_{c\bar s0}(2900)^{++}$ and $T^a_{c\bar s0}(2900)^{0}$ structures in the $DK$ and $D_{s} \pi$ final states in future experiments. The relative branching ratios between these decay channels will be very useful for understanding the nature of these states.

\begin{acknowledgments}
This work is supported by the National Natural Science Foundation of China under Grant No. 12175318, and No. 12075019, the National Key R$\&$D Program of China under Contracts No. 2020YFA0406400, the Natural Science Foundation of Guangdong Province of China under Grant No. 2022A1515011922, the Joint Large Scale Scientific Facility Funds of the National Natural Science Foundation of China (NSFC) and Chinese Academy of Sciences (CAS) under Con- tract No. U1932110. TGS is grateful for research funding from the Natural Sciences \& Engineering Research Council of Canada (NSERC).
\end{acknowledgments}

\newcommand{\noop}[1]{}

\end{document}